\documentclass[%
 reprint,
 amsmath,amssymb,
 aps,nofootinbib,   
]{revtex4-1}

\usepackage{bm}
\PassOptionsToPackage{linktocpage}{hyperref}
\usepackage[hyperindex,breaklinks]{hyperref}
\usepackage{enumitem}
\usepackage{slashed}

\usepackage{xcolor}
\usepackage{float}

\usepackage{array}
\usepackage{mathtools}

\usepackage{etoolbox}

\begin{document} 

\title{On the Wilsonian meaning of quantum error correction}

\author{  C\'esar G\'omez} 
\affiliation{Instituto de F\'{i}sica Te\'orica UAM-CSIC, Universidad Aut\'onoma de Madrid, Cantoblanco, 28049 Madrid, Spain}


\begin{abstract}
We sketch a recipe to define renormalization group transformations based on Kadanoff-Wilson block packing using a quantum error correction code. In such a case the RG transformations of the couplings are determined by the error matrix of the QEC code. In order to define the RG transformation of couplings we use Weinberg's sum rule for an error K\"{a}llen Lehmann function. We define an error beta function that for holographic AdS codes is conjectured to be zero. For holographic codes the relation between Weinberg's sum rule for the error K\"{a}llen Lehmann function and bulk locality is briefly discussed. 
 
 \end{abstract}
\maketitle


\section{Introduction}
The  Wilsonian definition of the renormalization group (RG)\cite{Wilsonphysrev} is a key ingredient of the modern approach to quantum field theory. The renormalization group first introduced as a formal transformation in \cite{SP} acquires its physical meaning in the study of QED at short distances first developed by Gell-Mann and Low in \cite{GL}.

The Wilsonian approach \cite{W,WK} promotes the RG into a practical tool to address the problem of solving the dynamics of systems having an enormously large number of degrees of freedom in a region with typical size given by the correlation length $\xi$. When we have many degrees of freedom in a region of size $\xi$ the relevant dynamics depends on {\it cooperative} phenomena and not so much on the microscopic characteristics of the Hamiltonian defining the local near neighbor interactions. The Wilsonian version of the RG defines, for a given dynamical system, a transformation $T$ on the space of local Hamiltonians that reduces, in each step, the effective number of degrees of freedom.

 The {\it cooperative} behavior is determined by the properties of the transformation $T$. In particular the {\it fixed point} of $T$ defines the critical behaviour, the corresponding scaling laws and the critical exponents.

An important step in the definition of the transformation $T$ was played by the idea of block packing introduced by Kadanoff \cite{rev,K}. To visualize the idea let us consider a spin lattice with lattice spacing given by $a$ and local near neighbor interactions. We can formally represent the Hamiltonian like $H=\sum K_{(i,i+1)} {\cal{O}}(\sigma_{i},\sigma_{i+1})$ where $i$ represents the lattice sites, $\sigma$ the spin operators and $K$ the couplings.

The block packing idea consists in defining a new lattice with lattice spacing $2a$. Each site of the new lattice represents a block of spins of the original lattice. The number of sites $b$ in each block will depend on the space dimension and the geometry of the lattice. The RG transformation defines new spin operators $\tilde \sigma_{\tilde i}$ for the new lattice sites and new couplings $\tilde K_{\tilde i,\tilde i +1}$. The transformed Hamiltonian $TH= \sum \tilde K {\cal{O}}(\tilde \sigma)$ acts on the block lattice configurations.

The RG group transformation for the couplings $\tilde K = T(K)$ allows us to identify the critical point as the fixed point $T(K^{*})=K^{*}$ as well as the critical exponents which are determined by $\frac{\mathrm{d}T}{\mathrm{d}K}$.

A crucial aspect of the RG transformation is to assume that {\it locality} is preserved meaning that the new spin variables $\tilde \sigma$ also interact through the same near neighbor type of local interaction. 

The original idea of Kadanoff was to assume that near criticality all the $b$ spins defining the block behave as a {\it unity} i.e. as a new spin variable. The ability of the RG transformation to extract the cooperative behavior of the system is partially grounded on the effective behavior of the block as a unity.

In more formal terms a RG transformation defines an embedding of the Hilbert space ${\cal{H}}$ describing the original spin lattice with, let us say, $n$ degrees of freedom in a region of size $\xi$ and dimension $2^n$, into a Hilbert space ${\cal{H}}_b$ with $\frac{n}{b}$ degrees of freedom:
\begin{equation}\label{RGmap}
T_{RG}: {\cal{H}} \rightarrow {\cal{H}}_b
\end{equation}

This formal map can be thought as a bundle {\it projection} where the {\it fiber} is defined as the set of spin configurations differing in {\it small} wave length fluctuations.

The local Hamiltonian $H$ acting on ${\cal{H}}$ is transformed into the local Hamiltonian $TH$ acting of ${\cal{H}}_b$. Moreover the local operator ${\cal{O}}(\sigma)$ acting on ${\cal{H}}$ is converted into the same local operator ${\cal{O}}(\tilde \sigma)$ acting on ${\cal{H}}_b$ for $\tilde \sigma$ representing the block spin operators.

The former description of the RG transformation strongly recalls the quantum information definition of quantum error correcting (QEC) codes \cite{QEC1,Steane,Got,Preskill,kribs,knill}.

QEC codes were designed as a practical tool to protect quantum information in transmission through a noisy channel. The idea is roughly as follows. We are interested in sending $k$ q-bits in such a way that errors created by a noisy channel can be identified and corrected. In order to do that we use a larger set, let us say $n$, of q-bits and we encode the Hilbert space ${\cal{H}}_k$ of the $k$ q-bits into the full Hilbert space ${\cal{H}}_n$ of the $n$ q-bits. This encoding protocol defines a map:
\begin{equation}
T_{QEC}: {\cal{H}}_k \rightarrow {\cal{H}}_n
\end{equation}
The image of ${\cal{H}}_k$ into ${\cal{H}}_n$ defines the {\it code subspace}. In order to implement the way potential errors, created by the noisy channel, can be corrected we introduce error operators $E_a$ acting on the large Hilbert space ${\cal{H}}_n$. The error operators can be characterized by a weight $w(E_a)$ which measures the number of sites on which the error operator acts non trivially. Let us now define the group ${\cal{E}}$ of error operators. Among this group we must identify the subset ${\cal{E}}_{c}$ of errors that can be corrected. This subset is characterized by the following two properties. For any element $|\psi\rangle$ in the code subspace let us define the {\it error orbit} as the set of states ${\cal{E}}_{c}|\psi\rangle$. 

To {\it identify} and distinguish the errors  we need to impose that for two orthogonal states in the code subspace $|\psi\rangle$ and $|\psi'\rangle$ the corresponding orbits ${\cal{E}}_{c}|\psi\rangle$ and ${\cal{E}}_{c}|\psi'\rangle$ are orthogonal. Moreover, in order to be able to {\it correct} the error without disturbing the quantum coherence of the encoded message state $|\psi\rangle$ we need to impose \cite{knill} that ${\cal{E}}_{c}$ acts on the code subspace as the identity operator i.e.
\begin{equation}\label{error}
\langle \psi|E|\psi\rangle = C(E)\text{ ,}
\end{equation}
for any operator in ${\cal{E}}_{c}$. We will refer to the matrix $C(E)$ defined in (\ref{error}) as the error matrix of the QEC code.

For a given QEC code we can define the {\it distance} $d$ as the maximal weight of error operators in ${\cal{E}}_{c}$. 

Once the set ${\cal{E}}_{c}$ of errors has been identified we can define, for a given QEC code, the {\it space of error orbits} ${\cal{H}}_{\cal{E}}$ as $\cup{\cal{E}}_{c}|\psi\rangle$ where the union is defined on the whole code subspace ${\cal{H}}_{\cal{C}}$. Moreover we can define the {\it projection} map
\begin{equation}
T_{QEC}^{\cal{E}} : {\cal{H}}_{{\cal{E}}} \rightarrow {\cal{H}}_{\cal{C}} \text{ .}
\end{equation}

This map can be thought as the analog of (\ref{RGmap}) for the following dictionary:

{\it Small fluctuations to be integrated} $\leftrightarrow$ {\it Errors to be corrected}

Next we shall use this formal analogy to define a RG transformation using a QEC code.\footnote{A recent approach to the RG that shares some similarities with the one described in this note uses tensor networks (see \cite{vidal} and references therein) to simulate the ground state wave function. By contrast our target will be to find explicit RG transformations on couplings using the properties of QEC codes.}

\section{QEC-RG}

Following Kadanoff's ideas we will consider a block of $b$ spins. Next we will use a QEC code $[b,1,d]$ where we use the $b$ spins in the block to encode one q-bit i.e. we use the QEC code to associate with a block just one spin operator $\tilde \sigma$. At this point we don't specify the distance $d$ of the QEC code. The spin sites in the block lattice are defined using the code basis $|\tilde 0\rangle$ and $|\tilde 1\rangle$ as
\begin{equation}
|\tilde 0\rangle = \sum c_j^{0} |j\rangle
\end{equation}
and 
\begin{equation}
|\tilde 1\rangle = \sum c_j^{1} |j\rangle \text{ ,}
\end{equation}
where the sum is over a basis of the unit block Hilbert space of dimension $2^b$. Generically the states $|\tilde 0\rangle$ and $|\tilde 1\rangle$ will be maximally entangled.

Now for each block spin configuration i.e. a state in ${\cal{H}}_b$ we define the corresponding state in ${\cal{H}}_n$ simply replacing the $|\tilde0\rangle$ and $|\tilde 1\rangle$ defining the spin sites of the state in ${\cal{H}}_b$ by the corresponding code representation defined above. Note that the former procedure can be {\it iterated} using the same fundamental QEC code $[b,1,d]$ reducing, in each step of the iteration, the effective correlation length.

Let us now consider a Hamiltonian $H=K{\cal{O}}(\sigma)$ for ${\cal{O}}(\sigma)$ a local operator defined using the original spin variables. The {\it renormalized} hamiltonian $TH= \tilde K {\cal{O}}(\tilde \sigma)$ for some renormalized coupling $\tilde K$ is by construction acting on the code subspace if we define $\tilde \sigma$ in terms of the QEC code spin states $|\tilde 0\rangle$ and $|\tilde 1\rangle$. In other words, ${\cal{O}}(\tilde \sigma)$ is by construction a {\it logical operator} of the QEC code, where by logical we mean that it maps the code subspace into itself (see for instance \cite{kribs,Preskill2}). In other words, if we define the block spin RG transformation $T$ using a QEC code then the transformed hamiltonian $TH$ is a logical operator. 

In order to identify the RG transformation on the couplings we need to discover $\tilde K$. 

Since the QEC code allows us to represent the states in the block spin lattice in terms of states in the original lattice Hilbert space we can act on a generic state of the block lattice i.e. a state in the code subspace, with the original Hamiltonian $K{\cal{O}}(\sigma)$. Generically the action of this Hamiltonian will {\it not} map the code subspace into itself. In other words, the original Hamiltonian is, in general, not a logical operator. Hence we introduce the following hypothesis.

{\it Hypothesis}: A QEC code defines a RG transformation of Hamiltonian $H=K{\cal{O}}(\sigma)$ if acting on a generic state $|\psi\rangle$ in the code subspace the state $H|\psi\rangle$ is in the {\it error orbit} of the state $K{\cal{O}}(\tilde \sigma)|\psi\rangle$ i.e.
\begin{equation}\label{main}
K{\cal{O}}( \sigma)|\psi\rangle \in {\cal{E}}_{c} (K{\cal{O}}(\tilde \sigma)|\psi\rangle)\text{ .}
\end{equation}

Now we can use the error matrix to define the amplitude $C(E;H)$ to create an error acting with the original Hamiltonian $H$ on the state in the code subspace. This allows us to define, for a given Hamiltonian, an {\it error K\"{a}llen Lehmann} function with support on the orbit  of errors,
\begin{equation}
\sigma_{KL}^{error}(E) = |C(E;H)|^2\text{ .}
\end{equation}
The next step will consist in using this error K\"{a}llen Lehmann function to extract the renormalization group transformation of the couplings
\begin{equation}\label{coupling}
\tilde K = N^{error} K\text{ .}
\end{equation}
In order to do it
we shall use the analog of Weinberg's sum rule, namely
\begin{equation}\label{sr}
1= Z^{error} + \int_{{\cal{E}}_c} |C(E;H)|^2\text{ ,}
\end{equation}
with $Z^{error} = |N^{error}|^2$. This assumes that the local operator ${\cal{O}}$ defining $H$ satisfies canonical commutation relations. Hence the renormalized hamiltonian acting on the code subspace is finally defined by

\begin{equation}
H = \tilde K {\cal{O}}(\tilde \sigma)\text{ ,}
\end{equation}
with
\begin{equation}
\tilde K = K\left ( 1- \int_{{\cal{E}}_c} |C(E;H)|^2\right )^{1/2}\text{ .}
\end{equation}

Physically we are visualizing the error matrix as setting  the {\it amplitude} to create an error $E$ using as noise the original hamiltonian $H(K)$ acting on a given state $|\psi\rangle$ in the code subspace ${\cal{H}}_{{\cal{C}}}$. Using an abuse of language we can denote this amplitude the {\it error  emission amplitude}.\footnote{A similar analogy for errors in the context of black hole physics was introduced in \cite{VV}.} 

The upper limit in the integration on errors of the {\it error K\"{a}llen Lehmann function} will be determined by the QEC code {\it distance}. It is important to notice that in each step of the iteration the effective number of errors on which we integrate the error K\"{a}llen Lehmann function increases. For instance the number of elementary one site error increases like $nb$ for $n$ the number of iterations. Thus the existence of a {\it continuum limit} depends on the convergency of the integral of $|c(E;H)|^2$ on the corresponding error orbit.

In case the QEC code admits a stabilizer group ${\cal{S}}$ \cite{Got} the renormalized Hamiltonian, in each step, is an element in $N({\cal{S}})-{\cal{S}}$ for $N({\cal{S}})$ the normalizer of ${\cal{S}}$ in ${\cal{E}}$. This constraints the minimal distance of a QEC code, that can define a RG transformation, by the weight of the operator Hamiltonian itself. Indeed since the renormalized Hamiltonian acts non trivially on the code subspace its weight must be bigger than the code distance.

As a final comment notice that the key point of condition (\ref{main}) lies in reducing the errors to those that can be {\it corrected}. This condition can be thought as the QEC analog of {\it renormalizability}. In fact the appearance in (\ref{main}) of generic operators in ${\cal{E}}$, for instance those in $N({\cal{S}})-{\cal{S}}$, will imply the generation in the renormalization process of new operators and consequently of new couplings.

In order to get a better intuition on the error K\"{a}llen Lehmann function it will be illustrative to recall the original analysis of RG in \cite{GL}. 

\section{Back to Gell-Mann and Low function}
The first example of RG equation for QED appears in \cite{GL}. Wilson's generalization in \cite{Wilsonphysrev} marks the beginning of the modern approach to RG in QFT. Using the notation of \cite{Wilsonphysrev} we have for QED
\begin{equation}\label{one}
e_{\lambda}^2 = e^2 d_c\left ( -\frac{\lambda^2}{m^2},e^2 \right )\text{ ,}
\end{equation}
with $\lambda$ representing the renormalization scale and with $d_c$ the renormalized quantum polarization of the photon.

GL function is defined as
\begin{equation}
\Psi\left(\frac{m^2}{\lambda^2}, e_{\lambda}\right) = \frac{\mathrm{d}e_{\lambda}^2}{\mathrm{d}(\ln \lambda)^2}\text{ .}
\end{equation}

The basic assumption of GL was the existence of the $m=0$ limit of $\Psi$. In this case the $\lambda=\infty$ limit of $e_{\lambda}$ is defined by the {\it fixed point} $\Psi(e_{\infty}^2,0)=0$. The crucial ingredient in the construction was to use the relation
\begin{equation}
e_{\lambda'}^2 = e_{\lambda}^2 \frac{d_c(\frac{\lambda'^2}{m^2},e^2)}{d_c(\frac{\lambda^2}{m^2},e^2)}\text{ .}
\end{equation}
That has a well defined $m=0$ limit provided $\lambda^2$ and $\lambda'^2$ are different from zero.

In practical terms this requires to define the $m=0$ limit of QED using a scale $\lambda^2 \neq 0$ that corresponds to off shell photons. The difficulties to implement this condition in the frame of the KLN theorem \cite{KLN} for canceling collinear IR divergences has been recently discussed in \cite{CR}.

The comment we want to make in this section is to stress the analogy between (\ref{one}) and the relation derived from the QEC discussion (\ref{coupling}).  The analog in perturbative QED  of $\int_{\cal{E}} |C(E)|^2$ is given by
$\ln\left (\frac{\lambda^2}{m^2}\right)$. Hence replacing $m$ by $\frac{1}{\xi}$ and the cutoff scale $\lambda$ by the maximal weight $w(E)$ in ${\cal{E}}_c$ i.e. the {\it dimension} $d(b)$ of the QEC code characterized by $b$  in units of the corresponding lattice spacing $a(b)$ we get as the QEC analog of QED 
\begin{equation}\label{two}
 Z^{error}  \sim 1- \ln\left(\frac{d(b)\xi}{a(b)}\right)
\end{equation}
This superficial analogy leads us to suggest that the continuum limit of a QEC code is not well defined if the probability of error emission is singular in the $\xi=\infty $ limit. 

Hence, as already indicated, the problem of UV completion in QEC code language could be interpreted as the convergency of the integral $\int |C(E;H)|^2$. 

Finally we can define an {\it error beta function} by
\begin{equation}
\beta^{error} \sim \frac{\partial Z^{error}}{\partial \tilde b}\text{ ,}
\end{equation}
where the standard variation with respect to the cutoff is replaced by the variation with respect to the size of the iterated block that we measure by $\tilde b=nb$.

\section{Holographic codes}
In the spirit of the AdS/CFT correspondence \cite{Mal} the notion of holographic QEC codes has been developed in \cite{AdS,Preskill2}. Next we will make some general comments in the frame of the holographic interpretation of the RG \cite{us}. 

Let us introduce an infrared parameter $\delta_{IR}$ to parametrize the region in bulk space at holographic coordinate $1-\delta_{IR}$ with $1$ representing the {\it boundary}.\footnote{This notation is based on the metric for AdS used in \cite{WS}} According to the holographic interpretation of the RG the corresponding QEC code should be defined in such a way that the Hilbert space ${\cal{H}}_{\delta_{IR}}$ describing the bulk region at coordinate $1-\delta_{IR}$ is a {\it code subspace} of the boundary Hilbert space ${\cal{H}}_{boundary}$ describing the boundary theory. To identify the corresponding QEC code we shall suggest the following hypothesis.

{\it Hypothesis}: The QEC code defining the bulk Hilbert space ${\cal{H}}_{\delta_{IR}}$ as a code subspace of the boundary Hilbert space, is a QEC code $[b(\delta_{IR}), 1, d]$ with
$b(\delta_{IR})$ determined by the UV/IR map \cite{WS}.\footnote{Note that the way $b$ increases with $\delta$ generically defines an hyperbolic geometry with negative curvature. In its discrete version the iterated application of the block QEC defines a sort of Cayley tree.}

In AdS this map is defined by looking for a region in the boundary of size $b$ such that the associated minimal surface in the bulk with boundary $b$ reach the point of coordinate $1-\delta_{IR}$. This hypothesis is the analog of the one in \cite{Preskill2} using the entanglement wedge. Moreover we expect in the special case of AdS/CFT geometry to have \cite{Preskill2}
\begin{equation}\label{scale}
d(b(\delta_{IR}))\sim b(\delta_{IR})\text{ .}
\end{equation}
Using as a guide the qualitative relation (\ref{two}) we can estimate the value of the {\it error beta function} that will be given by $\frac{\partial}{\partial b} (\frac{d(b)}{b^{1/d}})$ for $d$ the space dimension. So for those QEC codes satisfying (\ref{scale}) we get $\beta^{error} \sim 0$. In other words, holographic codes satisfying (\ref{scale}) i.e. AdS-codes, describe scale invariant dynamics. 

Let us finish with two general and purely qualitative comments. The former discussion motivates us to visualize the {\it holographic} description of a dynamical system defined by a Hamiltonian $H$ and a Hilbert space ${\cal{H}}$ as a one parameter family $(H_{\delta}, {\cal{H}}_{\delta})$ where ${\cal{H}}_{\delta'}$ is the code subspace of ${\cal{H}}_{\delta}$ for a QEC code $[b(\delta'-\delta),1,d(b(\delta'-\delta))]$ with $b(\delta)$ determined by a UV-IR map and with $d(b)$ scaling like $b$. 

The {\it gravity dual} dynamics is encoded, at the semiclassical level, in the two basic relations defining the QEC code, namely $b(\delta)$ and $d(b)$. In this frame the {\it quantum gravity} contribution to bulk correlators $\langle T( \phi(x,\delta), \phi(y,\delta))\rangle $ defined at the same value of the bulk coordinate ${\delta}$ is described by the contribution of errors in the intermediate states i.e. by the error K\"{a}llen Lehmann function. 

Note that now Weinberg's sum rule depends on the bulk locality properties of these correlators. Here the crucial point is to understand the contribution of errors (defining gravitational contributions to bulk correlators) that are not part of ${\cal{E}}_c$ i.e. errors that cannot be corrected. In very qualitative terms the limit of applicability of  scale invariant holographic QEC codes can be characterized by the inequality
\begin{equation}\label{ineq}
b(\delta) > d(b(\delta))\text{ .}
\end{equation}
Hence whenever we reach the regime in the bulk i.e. the coordinate $\delta$ for which we get the inequality (\ref{ineq}) errors that cannot be corrected start to dominate. As already discussed this sets the limit of what we can understand as {\it Wilsonian renormalizability}. Note that (\ref{ineq}) admits a purely semiclassical gravitational interpretation and very likely could be associated with the holographic version of dynamical generation of scales.

This last comment naturally leads to ask about the paradigmatic example of holography, namely a black hole. In this case the former picture will imply that the Hilbert space of the black hole of {\it age} $t$, let us call it ${\cal{H}}_t$, is a code subspace of the initial Hilbert space. The emitted radiation during this time will define a set of error operators \cite{VV}. In other words, after the time $t$ the BH state is represented in the {\it error bundle} by a couple $( |\psi\rangle (t), E(t))$ with $E(t)$ (the radiation) representing the point in the error fiber and $|\psi\rangle(t)$, representing the state of the BH at time $t$, in  ${\cal{H}}_t$ that plays the role of {\it base space}. 

Page's time \cite{Page} will be characterized by the moment where the errors associated to radiation are not acting as the unit matrix on the corresponding code space i.e. when the errors carry information on the BH quantum state or equivalently when we reach the regime characterized by (\ref{ineq}). We hope to address, in more quantitative terms, some of these issues in the future.

{\bf Acknowledgements.}
 This work was supported  by the ERC Advanced Grant 339169 "Selfcompletion" and the grant FPA2015-65480-P. We gratefully acknowledge the invitation from the Simons Center for Geometry and Physics, Stony Brook University where some parts of this research were completed.


\begin{thebibliography}{10}

\bibitem{Wilsonphysrev}
  K.~G.~Wilson,
  Phys.\ Rev.\ D {\bf 3} (1971) 1818.
  doi:10.1103/PhysRevD.3.1818


\bibitem{SP} E, C. G. Stueckelberg and A. Petermann, Helv. Phys. Acta 26, 499 (1953)
\bibitem{GL} M. Gell-Mann and F. E. Low, Phys. Rev. 95, 1300 (1954)

\bibitem{W}
  K.~G.~Wilson,
  Phys.\ Rev.\ B {\bf 4} (1971) 3174.
  doi:10.1103/PhysRevB.4.3174
\bibitem{WK}
  K.~G.~Wilson and J.~B.~Kogut,
  Phys.\ Rept.\  {\bf 12} (1974) 75.
  doi:10.1016/0370-1573(74)90023-4


\bibitem{rev}
  L.~P.~Kadanoff {\it et al.},
  Rev.\ Mod.\ Phys.\  {\bf 39} (1967) 395.
  doi:10.1103/RevModPhys.39.395


\bibitem{K}
  L.~P.~Kadanoff,
  Physics Physique Fizika {\bf 2} (1966) 263.
  doi:10.1103/PhysicsPhysiqueFizika.2.263


\bibitem{QEC1}
  A.~R.~Calderbank and P.~W.~Shor,
  Phys.\ Rev.\ A {\bf 54} (1996) 1098
  doi:10.1103/PhysRevA.54.1098
  [quant-ph/9512032].
 
 
 
  P.~W.~Shor,
  quant-ph/9605011.
 Shor, P. 1996  [quant-ph/9605011] 
 

  P.~W.~Shor and J.~A.~Smolin,
  quant-ph/9604006.
 Shor, P.  Smolin, J. 1996 [quant-ph/9604006] 
 \bibitem{Steane}
  A.~Steane,
  Proc.\ Roy.\ Soc.\ Lond.\ A {\bf 452} (1996) 2551
  doi:10.1098/rspa.1996.0136
  [quant-ph/9601029].
 
A.~Steane,
  
  Phys. Rev. Lett. 77, 793,(1996) 
  
  
  
\bibitem{Got}
  D.~Gottesman,
  quant-ph/9705052.

 D.~Gottesman,
  Phys.\ Rev.\ A {\bf 57} (1998) 127
  doi:10.1103/PhysRevA.57.127
  [quant-ph/9702029].


 D.~Gottesman,
  Phys.\ Rev.\ A {\bf 54} (1996) 1862
  doi:10.1103/PhysRevA.54.1862
  [quant-ph/9604038].


\bibitem{Preskill}
  J.~Preskill,
  Proc.\ Roy.\ Soc.\ Lond.\ A {\bf 454} (1998) 469
  doi:10.1098/rspa.1998.0171
  [quant-ph/9705032].
  

 J.~Preskill,
  Proc.\ Roy.\ Soc.\ Lond.\ A {\bf 454} (1998) 385
  doi:10.1098/rspa.1998.0167
  [quant-ph/9705031].
  	


\bibitem{kribs}
 D.W. Kribs, R. Laflamme,  and D. Poulin, Physical Review Letters 94, 180501 (2005).
 
 \bibitem{knill}
 E. Knill, R. Laflamme, Phys. Rev. A55, 900 (1997)
 
  C. Bennett, D. DiVincenzo, J. Smolin, and W. Wootters, Phys. Rev. A54, 3824 (1996).
  
\bibitem{Preskill2}
  F.~Pastawski and J.~Preskill,
  Phys.\ Rev.\ X {\bf 7} (2017) no.2,  021022
  doi:10.1103/PhysRevX.7.021022
  [arXiv:1612.00017 [quant-ph]].
 
  
  \bibitem{vidal} G. Evenbly and G. Vidal,  	Phys.Rev.Lett.115.180405 (2015) 	[cond-mat.str-el/1412.0732] 
   \bibitem{KLN}  T.  D.  Lee  and  M.  Nauenberg,  Phys. Rev.133(1964) B1549
    \bibitem{CR} C. Gomez and R. Letschka  [hep-th/19030131]
   \bibitem{Mal} J.M.  Maldacena,  Adv. Theor. Math. Phys.2, 231 (1998), [hep-th/9711200] 
   
    S.S.   Gubser, I.R. Klebanov and A.M. Polyakov,  Phys.Lett. B 428, 105 (1998), [hep-th/9802109] 
    
    E.   Witten,  Adv.Theor. Math. Phys. 2, 253 (1998), [hep-th/9802150]
     
    \bibitem{AdS}  A. Almheiri, X. Dong,  and D. Harlow, Journal of High Energy Physics 2015, 163 (2015), [hep-th/1411.7041] 
    
     F.  Pastawski,  B.  Yoshida,  D.  Harlow,    and  J.  Preskill,  Jour-nal  of  High  Energy  Physics 2015,  149  (2015),  [quant-ph/1503.06237]
     
     P. Hayden, S. Nezami, X.-L. Qi, N. Thomas, M. Walter,   and Z. Yang, Journal of High Energy Physics 2016, 9 (2016)
     
      	

F. Pastawski, B. Yoshida , D. Harlow, J. Preskill  JHEP 1506 (2015) 149
 \bibitem{VV} E. Verlinde, H. Verlinde JHEP 1310 (2013) 107
\bibitem{us}
  E. T. Akhmedov Phys. Lett.B442(1998) 152?158, [hep-th/9806217]  
  
  E. Alvarez and C. Gomez,Nucl. Phys.B541(1999) 441, [hep-th/9807226]  
  
  V. Balasubramanian and P. Kraus,Phys. Rev. Lett.83(1999) 3605, [hep-th/9903190]
  
  K. Skenderis and P. K. Townsend,
    Phys. Lett.B468(1999) 46?51, [hep-th/9909070]
    
    J. de Boer, E. P. Verlinde and H. L. Verlinde,JHEP08(2000) 003, [hep-th/9912012].

 \bibitem{WS} 
L. Susskind,  E. Witten  [hep-th/9805114]
 \bibitem{Page} D.N.   Page,  Phys.  Rev.  Lett.  71(1993) 3743-3746; [hep-th/9306083]
\end{thebibliography}
\end{document}